\documentclass[12pt]{article}
\usepackage{graphicx}
\usepackage{epsfig}
\usepackage{epstopdf}
\DeclareGraphicsExtensions{.pdf,.eps,.png,.jpg,.mps}
\usepackage{caption}
\usepackage{float}

\usepackage{cite}

\def\lsim{\mathrel{\rlap{\lower3pt\hbox{\hskip0pt$\sim$}}
     \raise1pt\hbox{$<$}}}         
\def\gsim{\mathrel{\rlap{\lower4pt\hbox{\hskip1pt$\sim$}}
     \raise1pt\hbox{$>$}}}         

\usepackage{amsmath}
\usepackage{amsfonts}

\begin{document}
\begin{titlepage}

\centerline{\Large \bf Statistical Risk Models}
\medskip

\centerline{Zura Kakushadze$^\S$$^\dag$\footnote{\, Zura Kakushadze, Ph.D., is the President of Quantigic$^\circledR$ Solutions LLC,
and a Full Professor at Free University of Tbilisi. Email: zura@quantigic.com} and Willie Yu$^\sharp$\footnote{\, Willie Yu, Ph.D., is a Research Fellow at Duke-NUS Medical School. Email: willie.yu@duke-nus.edu.sg}}
\bigskip

\centerline{\em $^\S$ Quantigic$^\circledR$ Solutions LLC}
\centerline{\em 1127 High Ridge Road \#135, Stamford, CT 06905\,\,\footnote{\, DISCLAIMER: This address is used by the corresponding author for no
purpose other than to indicate his professional affiliation as is customary in
publications. In particular, the contents of this paper
are not intended as an investment, legal, tax or any other such advice,
and in no way represent views of Quantigic$^\circledR$ Solutions LLC,
the website \underline{www.quantigic.com} or any of their other affiliates.
}}
\centerline{\em $^\dag$ Free University of Tbilisi, Business School \& School of Physics}
\centerline{\em 240, David Agmashenebeli Alley, Tbilisi, 0159, Georgia}
\centerline{\em $^\sharp$ Centre for Computational Biology, Duke-NUS Medical School}
\centerline{\em 8 College Road, Singapore 169857}
\medskip
\centerline{(February 14, 2016)}

\bigskip
\medskip

\begin{abstract}
{}We give complete algorithms and source code for constructing statistical risk models, including methods for fixing the number of risk factors. One such method is based on eRank (effective rank) and yields results similar to (and further validates) the method of \cite{Het}. We also give a complete algorithm and source code for computing eigenvectors and eigenvalues of a sample covariance matrix which requires i) no costly iterations and ii) the number of operations linear in the number of returns. The presentation is intended to be pedagogical and oriented toward practical applications.
\end{abstract}
\medskip
\end{titlepage}

\newpage

\section{Introduction}

{}Multifactor risk models\footnote{\,
For a partial list of related works, see
\cite{Q1}, \cite{Q2}, \cite{Q3}, \cite{Q4}, \cite{Q5}, \cite{Q6}, \cite{Q7}, \cite{Q8}, \cite{Q9}, \cite{Q10},
\cite{Q11}, \cite{Q12}, \cite{Q13}, \cite{Q14}, \cite{Q15}, \cite{Q16}, \cite{Q17}, \cite{Q18}, \cite{Q19}, \cite{Q20},
\cite{Q21}, \cite{Q22}, \cite{Q23}, \cite{Q24}, \cite{Q26, Q25}, \cite{Q27}, \cite{Q28}, \cite{Q29, Q30},
\cite{Q31, Q32, Q33}, \cite{Q34}, \cite{Q35}, \cite{Q36}, \cite{Q37, Q38, Q39, Q40},
\cite{Q41}, \cite{Q42, Q43}, \cite{Q44, Q45}, \cite{Q46}, \cite{Q47}, \cite{Q48}, \cite{Q49}, \cite{Q50},
\cite{Q51}, \cite{Q52}, \cite{Q53}, \cite{Q54}, \cite{Q55, Q56}, (Kakushadze 2015a,b,c,d, 2016), \cite{CustomRM}, \cite{HetPlus},
\cite{Q62}, \cite{Q63}, \cite{Q64}, \cite{Q65}, \cite{Q66}, \cite{Q67}, \cite{Q68}, \cite{Q69}, \cite{Q70},
\cite{Q71}, \cite{Q72}, \cite{Q73}, \cite{Q74, Q75}, \cite{Q76}, \cite{Q77}, \cite{Q78}, \cite{Q79}, \cite{Q80},
\cite{Q81}, \cite{Q82}, \cite{Q83}, \cite{Q84}, \cite{Q85, Q86, Q87}, \cite{Q88}, \cite{Q89}, \cite{Q90, Q91}, \cite{Q92},
\cite{Q93, Q94}, \cite{Q95, Q96}, \cite{Q97}, \cite{Q98}, \cite{Q99}, \cite{Q100}, \cite{Q101}, \cite{Q102},
and references therein.
}
are a popular risk management tool, e.g., in portfolio optimization. For stock portfolios, in their most popular incarnations, multifactor risk models are usually constructed based on industry and style risk factors.\footnote{\, Shortcomings with traditional implementations are detailed in \cite{HetPlus}.} However, in some cases such constructions are unavailable, e.g., because any industry classification (or similar) is lacking, any relevant style factors are impossible to define, etc. In fact, this is generally the case when the underlying returns are not for equities but some other ``instruments", e.g., quantitative trading alphas (expected returns).

{}In such cases one usually resorts to statistical risk models. Often times these are thought of in the context of principal components of a sample covariance (or correlation) matrix of returns. More generally, one can think of statistical risk models as constructed solely based on the time series of the underlying returns and no additional information. The purpose of these notes is to provide a simple and pedagogical discussion of statistical risk models oriented toward practical applications.

{}In Section \ref{sec.2} we set up our discussion by discussing the sample covariance matrix, generalities of factor models, the requirement that factor models reproduce in-sample variances, and how a $K$-factor statistical risk model can be simply constructed by starting from the sample covariance (or correlation) matrix, writing down its spectral representation via principal components, truncating the sum by keeping only  the first $K$ principal components, and compensating for the deficit in the variances (i.e., on the diagonal of the resultant matrix) by adding specific (idiosyncratic) risk. This (generally)\footnote{\, Assuming no two returns are 100\% pair-wise (anti-)correlated.} results in a positive-definite (and thus invertible) risk model covariance matrix so long as $K < M$, where $M+1$ is the number of observations in the time series. This holds even if $M < N$, in which case the sample covariance matrix is singular. In fact, one of the main motivations for considering factor models in the first instance is that in most practical applications $M <N$ (and often $M \ll N$), and even if $M\geq N$, in which case the sample covariance matrix is nonsingular, it is still out-of-sample unstable unless $M\gg N$, which is seldom (if ever) the case in practice. Factor models are intended to reduce this instability to a degree.\footnote{\, Albeit in the case of statistical risk models, which are based on the very same returns used in computing the sample covariance matrix, this instability sizably seeps into the factor model.}

{}The beauty of the statistical risk model construction is its simplicity. However, one must fix the number of risk factors $K$. We discuss two simple methods for fixing $K$ in Section \ref{sub.fix.K} (with variations). One is that of \cite{Het}. Another, very different looking method, is based on our adaptation of eRank (effective rank) of \cite{RV} and yields results similar to (and further validates) that of \cite{Het}. We use intraday alphas of \cite{MeanRev} and backtest these methods out-of-sample. The method of \cite{Het} backtests better. We give R source code for computing a $K$-factor statistical risk model with $K$ fixed via the aforementioned two methods (with variations) in Appendix \ref{app.A}.\footnote{\, The source code given in Appendix \ref{app.A}, Appendix \ref{app.B} and Appendix \ref{app.C} is not written to be ``fancy" or optimized for speed or in any other way. Its sole purpose is to illustrate the algorithms described in the main text in a simple-to-understand fashion. See Appendix \ref{app.D} for some legalese.}

{}In Section \ref{sub.pc} we discuss how to compute principal components based on the returns. The ``na\"ive" method is the power iterations method, which is applicable to more general matrices. However, it requires iterations and is computationally costly.\footnote{\, We give R code for this method in Appendix \ref{app.B}.} Because here we are dealing with sample covariance matrices, there is a simpler and faster way of computing principal components when $M\ll N$ that does not require any costly iterations and involves only ${\cal O}(M^2 N)$ operations. We discuss this method in detail in Section \ref{sub.pc} and give R source code for it in Appendix \ref{app.C}. The main purpose of this exercise is to set up our further discussion in Section \ref{sub.pc}, where we explain that statistical risk models are simply certain deformations of the sample covariance matrix. We then also discuss ``nontraditional" statistical risk models such as shrinkage \cite{LW}, which are also deformations of the sample covariance matrix, but involve $M$ principal components as opposed to $K < M$ principal components. Generally, ``nontraditional" models underperform.

{}We then take this a step further and explain that optimization using a statistical risk model is well-approximated by a weighted regression, where the regression is over the factor loadings matrix (i.e., the $K$ principal components), and the weights are inverse specific variances. More precisely, this holds when the number of underlying returns $N\gg 1$, which is the case in most applications. In fact, optimization reduces to a weighted regression for $N\gg 1$ in a wider class of risk models that lack any ``clustering" structure (we clarify the meaning of this statement in Section \ref{sub.pc}).

{}We briefly conclude in Section \ref{sec.conc}, where we discuss additional backtests, etc.

\section{Statistical Risk Models}\label{sec.2}
\subsection{Sample Covariance Matrix}

{}So, we have $N$ instruments (e.g., stocks) with the time series of returns. Each time series contains $M+1$ observations corresponding to times $t_s$, and we will denote our returns as $R_{is}$, where $i=1,\dots,N$ and $s=1,\dots,M,M+1$ ($t_1$ is the most recent observation). The sample covariance matrix (SCM) is given by\footnote{\, The difference between the unbiased estimate with $M$ in the denominator vs. the maximum likelihood estimate with $M+1$ in the denominator is immaterial; in most applications $M \gg 1$.}
\begin{equation}\label{sample.cov.mat}
 C_{ij} = {1\over M}\sum_{s=1}^{M+1} X_{is}~X_{js}
\end{equation}
where $X_{is} = R_{is} - {\overline R}_i$ are serially demeaned returns; ${\overline R}_i = {1\over {M+1}}\sum_{s=1}^{M+1} R_{is}$.

{}We are interested in cases where $M < N$, in fact, $M\ll N$. When $M < N$, $C_{ij}$ is singular: we have $\sum_{s=1}^{M+1} X_{is} = 0$, so only $M$ columns of the matrix $X_{is}$ are linearly independent. Let us eliminate the last column: $X_{i,M+1}=-\sum_{s=1}^M X_{is}$. Then we can express $C_{ij}$ via the first $M$ columns:
\begin{equation}\label{SCM}
 C_{ij} = \sum_{s,s^\prime=1}^{M} X_{is}~\phi_{ss^\prime}~X_{js^\prime}
\end{equation}
Here $\phi_{ss^\prime} = \left(\delta_{ss^\prime} + u_s u_{s^\prime}\right)/M$ is a nonsingular $M\times M$ matrix ($s,s^\prime = 1,\dots,M$); $u_s \equiv 1$ is a unit $M$-vector. Note that $\phi_{ss^\prime}$ is a 1-factor model (see below).

{}So, when $M<N$, the sample covariance matrix $C_{ij}$ is singular with $M$ nonzero eigenvalues. In this case we cannot invert $C_{ij}$, which is required in, e.g., optimization (mean-variance optimization \cite{Q74}, Sharpe ratio maximization \cite{Sharpe94}, etc.). Furthermore, unless $M\gg N$, which is almost never (if ever) the case in practical applications, the off-diagonal elements of $C_{ij}$ (covariances) generally are not expected to be stable out-of-sample. In contrast, the diagonal elements (variances) typically are much more stable out-of-sample and can be relatively reliably computed even for $M\ll N$ (which, in fact, is often the case in practical applications). So, we need to replace the sample covariance matrix $C_{ij}$ by another {\em constructed} matrix -- call it $\Gamma_{ij}$ -- that is much more stable out-of-sample and invertible (positive-definite). That is, we must build a risk model.

\subsection{Factor Models}

{}A popular method -- at least in the case of equities -- for constructing a nonsingular replacement $\Gamma_{ij}$ for $C_{ij}$ is via a factor model:
\begin{equation}\label{fac.mod}
 \Gamma_{ij} = \xi_i^2~\delta_{ij} + \sum_{A,B=1}^K \Omega_{iA}~\Phi_{AB}~\Omega_{jB}
\end{equation}
Here: $\xi_i$ is the specific (a.k.a. idiosyncratic) risk for each return; $\Omega_{iA}$ is an $N\times K$ factor loadings matrix; and $\Phi_{AB}$ is a $K\times K$ factor covariance matrix (FCM), $A,B=1,\dots,K$. The number of factors $K\ll N$ to have FCM more stable than SCM. And $\Gamma_{ij}$ is positive-definite (and invertible) if FCM is positive-definite and all $\xi_i^2 > 0$.

\subsection{Total Variances}

{}The main objective of a risk model is to predict the covariance matrix out-of-sample as precisely as possible, including the out-of-sample variances. However, even though this requirement is often overlooked in practical applications, a well-built factor model had better reproduce the in-sample variances. That is, we require that the risk model variances $\Gamma_{ii}$ be equal the in-sample variances $C_{ii}$:
\begin{equation}\label{tot.risk}
 \Gamma_{ii} = C_{ii} = \sigma_i^2
\end{equation}
Furthermore, as mentioned above, the $N$ variances $C_{ii}$ are relatively stable out-of-sample. It is therefore the $N(N-1)$ off-diagonal covariances, which are generally unstable out-of-sample, we must actually model. Put differently, we must model the {\em correlations} $\Psi_{ij}$, $i\neq j$, where $\Psi_{ij}= C_{ij}/\sigma_i\sigma_j$ is the sample correlation matrix, whose diagonal elements  $\Psi_{ii} \equiv 1$. So, we need to replace the sample correlation matrix by another constructed matrix -- let us call it ${\widetilde \Gamma}_{ij}$ -- that is much more stable out-of-sample and invertible (positive-definite) subject to the conditions
\begin{equation}\label{diag.psi}
 {\widetilde \Gamma}_{ii} = \Psi_{ii} \equiv 1
\end{equation}
Once we build ${\widetilde\Gamma}_{ij}$, the risk model covariance matrix is given by $\Gamma_{ij} = \sigma_i\sigma_j{\widetilde \Gamma}_{ij}$. The advantage of modeling the correlation matrix $\Psi_{ij}$ via ${\widetilde\Gamma}_{ij}$ as opposed to modeling the covariance matrix $C_{ij}$ by $\Gamma_{ij}$ is that the sample variances $C_{ii} = \sigma_i^2$ have a highly skewed (quasi log-normal) distribution, while $\Psi_{ii}$ are uniform. In the following, in the main text\footnote{\, However, in Appendix \ref{app.A} we give the source code which has an option to compute the factor model directly for $C_{ij}$ as opposed to via $\Psi_{ij}$.}  we will always focus on modeling $\Psi_{ij}$ and for the sake of notational simplicity we will omit the twiddle on ${\widetilde\Gamma}_{ij}$, i.e., we will model
\begin{equation}\label{CorMat}
 \Psi_{ij} = \sum_{s,s^\prime=1}^{M} Y_{is}~\phi_{ss^\prime}~Y_{js^\prime}
\end{equation}
via (\ref{fac.mod}) subject to $\Gamma_{ii}\equiv 1$. Here $Y_{is} = X_{is}/\sigma_i$.

\subsection{Principal Components}

{}Looking at (\ref{CorMat}), it resembles a factor model, except that it has no specific risk (so it is singular). We cannot simply add some specific risk ad hoc to (\ref{CorMat}) as this would violate the requirement that $\Gamma_{ii}\equiv 1$ (as the resulting $\Gamma_{ii}$ would be greater than 1). Therefore, to add some specific risk, we must simultaneously reduce the diagonal contribution from the factor risk. All values of $s,s^\prime = 1,\dots,M$ in the sum in (\ref{CorMat}) enter on the equal footing, in fact, we have a full ${\bf Z}_M$ permutational symmetry under which $s\rightarrow s + 1$, $s^\prime\rightarrow s^\prime + 1$, and $s\,(s^\prime) > M$ is identified with $s - M$ ($s^\prime - M)$. When reducing factor risk we must either preserve this symmetry or somehow break it.

{}In fact, we can choose a different -- but equivalent -- basis, where this symmetry is not explicit and it is more evident how to ``trim" the factor risk. Let $V_i^{(a)}$, $a=1,\dots,N$, be the principal components of $\Psi_{ij}$ forming an orthonormal basis
\begin{eqnarray}\label{Psi.eigen}
 &&\sum_{j=1}^N \Psi_{ij}~V_j^{(a)} = \lambda^{(a)}~V_i^{(a)}\\
 &&\sum_{i=1}^N V_i^{(a)}~V_i^{(b)} = \delta_{ab}
\end{eqnarray}
such that the eigenvalues $\lambda^{(a)}$ are ordered decreasingly: $\lambda^{(1)} > \lambda^{(2)} >\dots$. More precisely, some eigenvalues may be degenerate. For simplicity -- and this is not critical here -- we will assume that all positive eigenvalues are non-degenerate. However, we can have multiple null eigenvalues. Typically, the number of nonvanishing eigenvalues\footnote{\, This number can be smaller if some returns are 100\% correlated or anti-correlated. For the sake of simplicity -- and this not critical here -- we will assume that there are no such returns.} is $M$, where, as above, $M+1$ is the number of observations in the return time series. So, we have
\begin{equation}\label{CM.PC}
 \Psi_{ij} = \sum_{a = 1}^M V_i^{(a)}~\lambda^{(a)}~V_j^{(a)}
\end{equation}
This again resembles a factor model (with a diagonal factor covariance matrix). However, the ${\bf Z}_M$ symmetry is gone and we can readily ``trim" the factor risk.

{}This is simply done by keeping only $K < M$ first principal components in the sum in (\ref{CM.PC}) and replacing the diagonal contribution of the dropped $M-K$ principal components via the specific risk:
\begin{eqnarray}\label{PC}
 && \Gamma_{ij} = {\xi}_i^2~\delta_{ij} + \sum_{A=1}^K \lambda^{(A)}~V_i^{(A)}~V_j^{(A)}\\
 && {\xi}_i^2 = 1 - \sum_{A=1}^K \lambda^{(A)}\left(V_i^{(A)}\right)^2\label{PC.xi}
\end{eqnarray}
This corresponds to taking the factor loadings matrix and factor covariance matrix of the form
\begin{eqnarray}\label{FLM.PC}
 &&{\Omega}_{iA} = \sqrt{\lambda^{(A)}}~V_i^{(A)},~~~A=1,\dots,K\\
 &&\Phi_{AB} = \delta_{AB}\label{CorMat.PC}
\end{eqnarray}
This construction is nicely simple. However, what should $K$ be?

\section{Fixing Factor Number}\label{sub.fix.K}

{}When $K = M$ we have $\Gamma_{ij} = \Psi_{ij}$, which is singular. Therefore, we must have $K \leq K_{max} < M$. So, what is $K_{max}$? And what is $K_{min}$ (other than the evident $K_{min} = 1$)? It might be tempting to do complicated and convoluted things. We will not do this here. Instead, we will follow a pragmatic approach. One simple (``minimization" based) algorithm was set forth in \cite{Het}. We review it below and then give yet another simple algorithm based on eRank (effective rank).

\subsection{``Minimization" Algorithm}

{}The idea is simple \cite{Het}. It is based on the observation that, as $K$ approaches $M$, $\mbox{min}({\xi}^2_i)$ goes to 0 (i.e., less and less of the total risk is attributed to the specific risk, and more and more of it is attributed to the factor risk), while as $K$ approaches 0, $\mbox{max}({\xi}^2_i)$ goes to 1 (i.e., less and less of the total risk is attributed to the factor risk, and more and more of it is attributed to the specific risk). So, we can define $K$ as follows:
\begin{eqnarray}\label{K}
 &&|g(K) - 1| \rightarrow \mbox{min}\\
 \label{g}
 &&g(K) = \sqrt{\mbox{min}({\xi}^2_i)} + \sqrt{\mbox{max}({\xi}^2_i)}
\end{eqnarray}
This simple algorithm works pretty well in practical applications.\footnote{\, The distribution of ${\xi}^2_i$ is skewed; typically, ${\xi}^2_i$ has a tail at higher values, while $\ln({\xi}^2_i)$ has a tail at lower values, and the distribution is only roughly log-normal. So $K$ is not (the floor/cap of) $M/2$, but somewhat higher, albeit close to it. See \cite{Het} for an illustrative example. \label{fn.xi.distrib}}

\subsection{Effective Rank}

{}Another simple method is to set (here $\mbox{Round}(\cdot)$ can be replaced by $\mbox{floor}(\cdot) = \lfloor\cdot\rfloor$)
\begin{equation}\label{eq.eRank}
 K = \mbox{Round}(\mbox{eRank}(\Psi))
\end{equation}
Here $\mbox{eRank}(Z)$ is the effective rank \cite{RV} of a symmetric semi-positive-definite (which suffices for our purposes here) matrix $Z$. It is defined as
\begin{eqnarray}
 &&\mbox{eRank}(Z) = \exp(H)\\
 &&H = -\sum_{a=1}^L p_a~\ln(p_a)\\
 &&p_a = {\lambda^{(a)} \over \sum_{b=1}^L \lambda^{(b)}}
\end{eqnarray}
where $\lambda^{(a)}$ are the $L$ {\em positive} eigenvalues of $Z$, and $H$ has the meaning of the (Shannon a.k.a. spectral) entropy \cite{Campbell60}, \cite{YGH}.

{}The meaning of $\mbox{eRank}(Z)$ is that it is a measure of the effective dimensionality of the matrix $Z$, which is not necessarily the same as the number $L$ of its positive eigenvalues, but often is lower. This is due to the fact that many returns can be highly correlated (which manifests itself by a large gap in the eigenvalues) thereby further reducing the effective dimensionality of the correlation matrix.

\subsection{A Variation}\label{sub.k.prime}

{}When the average correlation\footnote{\, Instead we can define ${\overline \Psi} = {1\over N(N-1)}\sum_{i,j=1;~i\neq j}^N$. Since $N\gg 1$, the difference is immaterial.} ${\overline \Psi} = {1\over N^2}\sum_{i,j=1}^N\Psi_{ij}$ is high, then both the ``minimization" and eRank based algorithms can produce low values of $K$ (including 1). This is because in this case $\lambda^{(1)}\gg 1$ and there is a large gap in the eigenvalues. To circumvent this, we can define $K = K^\prime + 1$, where $K^\prime$ is defined as above via the ``minimization" or eRank based algorithms for the matrix
\begin{equation}
 \Psi_{ij}^\prime = \sum_{a = 2}^M V_i^{(a)}~\lambda^{(a)}~V_j^{(a)}
\end{equation}
I.e., we simply drop the first eigenvalue, determine the corresponding value of $K^\prime$, and add 1 to it. Appendix \ref{app.A} provides R source code for both the ``minimization" and eRank based algorithms with and without utilizing the $K^\prime$ based definition.

\subsection{Some Backtests}\label{sub.backtests}

{}Let us backtest the above algorithms for fixing $K$ via utilizing the same backtesting procedure as in \cite{Het}. The remainder of this subsection very closely follows most parts of Section 6 of \cite{Het}.\footnote{\, We ``rehash" it here not to be repetitive but so that the presentation herein is self-contained.}

\subsubsection{Notations}

{}Let $P_{is}$ be the time series of stock prices, where $i=1,\dots,N$ labels the stocks, and $s=1,\dots,M+1$ labels the trading dates, with $s=1$ corresponding to the most recent date in the time series. The superscripts $O$ and $C$ (unadjusted open and close prices) and $AO$ and $AC$ (open and close prices fully adjusted for splits and dividends) will distinguish the corresponding prices, so, e.g., $P^C_{is}$ is the unadjusted close price. $V_{is}$ is the unadjusted daily volume (in shares). Also, for each date $s$ we define the overnight return as the previous-close-to-open return:
\begin{equation}\label{c2o.ret}
 E_{is} = \ln\left({P^{AO}_{is} / P^{AC}_{i,s+1}}\right)
\end{equation}
This return will be used in the definition of the expected return in our mean-reversion alpha. We will also need the close-to-close return
\begin{equation}\label{c2c.ret}
 R_{is} = \ln\left({P^{AC}_{is} / P^{AC}_{i,s+1}}\right)
\end{equation}
An out-of-sample (see below) time series of these returns will be used in constructing the risk models. All prices in the definitions of $E_{is}$ and $R_{is}$ are fully adjusted.

{}We assume that: i) the portfolio is established at the open\footnote{\, This is a so-called ``delay-0" alpha: the same price, $P^O_{is}$ (or adjusted $P^{AO}_{is}$), is used in computing the expected return (via $E_{is}$) and as the establishing fill price.} with fills at the open prices $P^O_{is}$; ii) it is liquidated at the close on the same day -- so this is a purely intraday alpha -- with fills at the close prices $P^C_{is}$; and iii) there are no transaction costs or slippage -- our aim here is not to build a realistic trading strategy, but to test {\rm relative} performance of various risk models and see what adds value to the alpha and what does not. The P\&L for each stock
\begin{equation}
 \Pi_{is} = H_{is}\left[{P^C_{is}\over P^O_{is}}-1\right]
\end{equation}
where $H_{is}$ are the {\em dollar} holdings. The shares bought plus sold (establishing plus liquidating trades) for each stock on each day are computed via $Q_{is} = 2 |H_{is}| / P^O_{is}$.

\subsubsection{Universe Selection}\label{sub.univ}

{}For the sake of simplicity,\footnote{\, In practical applications, the trading universe of liquid stocks typically is selected based on market cap, liquidity (ADDV), price and other (proprietary) criteria.} we select our universe based on the average daily dollar volume (ADDV) defined via (note that $A_{is}$ is out-of-sample for each date $s$):
\begin{equation}\label{ADDV}
 A_{is}= {1\over d} \sum_{r=1}^d V_{i, s+r}~P^C_{i, s+r}
\end{equation}
We take $d=21$ (i.e., one month), and then take our universe to be the top 2000 tickers by ADDV. To ensure that we do not inadvertently introduce a universe selection bias, we rebalance monthly (every 21 trading days, to be precise). I.e., we break our 5-year backtest period (see below) into 21-day intervals, we compute the universe using ADDV (which, in turn, is computed based on the 21-day period immediately preceding such interval), and use this universe during the entire such interval. We do have the survivorship bias as we take the data for the universe of tickers as of 9/6/2014 that have historical pricing data on http://finance.yahoo.com (accessed on 9/6/2014) for the period 8/1/2008 through 9/5/2014. We restrict this universe to include only U.S. listed common stocks and class shares (no OTCs, preferred shares, etc.) with BICS (Bloomberg Industry Classification System) sector assignments as of 9/6/2014.\footnote{\, The choice of the backtesting window is intentionally taken to be exactly the same as in \cite{Het} to simplify various comparisons, which include the results therefrom.} However, as discussed in detail in Section 7 of \cite{MeanRev}, the survivorship bias is not a leading effect in such backtests.\footnote{\, Here we are after the {\em relative outperformance}, and it is reasonable to assume that, to the leading order, individual performances are affected by the survivorship bias approximately equally as the construction of all alphas and risk models is ``statistical" and oblivious to the universe.}

\subsubsection{Backtesting}\label{sub.back}

{}We run our simulations over a period of 5 years (more precisely, 1260 trading days going back from 9/5/2014, inclusive). The annualized return-on-capital (ROC) is computed as the average daily P\&L divided by the intraday investment level $I$ (with no leverage) and multiplied by 252. The annualized Sharpe Ratio (SR) is computed as the daily Sharpe ratio multiplied by $\sqrt{252}$. Cents-per-share (CPS) is computed as the total P\&L divided by the total shares traded.\footnote{\, As mentioned above, we assume no transaction costs, which are expected to reduce the ROC of the optimized alphas by the same amount as all strategies trade the exact same amount by design. Therefore, including the transaction costs would have no effect on the actual {\em relative outperformance} in the horse race, which is what we are after here.}

\subsubsection{Optimized Alphas}\label{sub.opt}

{}The optimized alphas are based on the expected returns $E_{is}$ optimized via Sharpe ratio maximization using the risk models we are testing, i.e., the covariance matrix ${\widehat\Gamma}_{ij} = \sigma_i\sigma_j\Gamma_{ij}$ with $\Gamma_{ij}$ given by (\ref{PC}), which we compute every 21 trading days (same as for the universe). For each date (we omit the index $s$) we maximize the Sharpe ratio subject to the dollar neutrality constraint:
\begin{eqnarray}
 &&{\cal S} = {\sum_{i=1}^N H_i~E_i\over{\sqrt{\sum_{i,j=1}^N {\widehat \Gamma}_{ij}~H_i~H_j}}} \rightarrow \mbox{max}\\
 &&\sum_{i=1}^N H_i = 0\label{d.n.opt}
\end{eqnarray}
The solution is given by
\begin{equation}\label{H.opt}
 H_i = -\eta \left[\sum_{j = 1}^N {\widehat \Gamma}^{-1}_{ij}~E_j - \sum_{j=1}^N {\widehat \Gamma}^{-1}_{ij}~{{\sum_{k,l=1}^N {\widehat \Gamma}^{-1}_{kl}~E_l}\over{\sum_{k,l = 1}^N {\widehat \Gamma}^{-1}_{kl}}}\right]
\end{equation}
where ${\widehat \Gamma}^{-1}$ is the inverse of ${\widehat \Gamma}$, and $\eta > 0$ (mean-reversion alpha) is fixed via (we set the investment level $I$ to \$20M in our backtests)
\begin{equation}
 \sum_{i=1}^N \left|H_i\right| = I
\end{equation}
Note that (\ref{H.opt}) satisfies the dollar neutrality constraint (\ref{d.n.opt}).

{}The simulation results are given in Table \ref{table1} for $K$ obtained via the ``minimization" and eRank based algorithms with and without utilizing the $K^\prime$ based definition (see Subsection \ref{sub.k.prime}) with $\mbox{Round}(\cdot)$ and $\mbox{floor}(\cdot)$ in (\ref{eq.eRank}). The ``minimization" and eRank methods not based on $K^\prime$ produce similar results, which further validates the ``minimization" method of \cite{Het}. The slight improvement in CPS in the eRank method is immaterial and disappears when we impose position bounds (which in this case are the same as trading bounds as the strategy is purely intraday)
\begin{equation}\label{liq}
 |H_{is}| \leq 0.01~A_{is}
\end{equation}
where $A_{is}$ is ADDV defined in (\ref{ADDV}). These backtests use the R code in Appendix C of \cite{Het} for optimization with bounds. Table \ref{table2} gives the simulation results with these bounds for all of the above cases except for the eRank method with the $K^\prime$ based definition. In the latter case, because the typical value of $K$ is close to or the same as the maximum allowed $K_{max} = M - 1 = 19$ (see rows 5 and 6 in Table \ref{table3}), some of the desired holdings come out to be large compared with the (reasonable) bounds (\ref{liq}) (see row 7 in Table \ref{table3}). The ``average correlation" ${\overline\Psi}$ is not very high (nor is it very low -- see row 8 in Table \ref{table3}), so it is just as well that the value of $K$ is not low and the $K^\prime$ based method is overkill, which is why it underperforms.

\section{Principal Components, Deformations, etc.}\label{sub.pc}
\subsection{Power Iterations}

{}To construct the statistical risk models, we need to compute the first $M$ principal components of $\Psi_{ij}$. One way is to successively use the power iterations method \cite{PowerIt}, which usually costs more than ${\cal O}(M^2 N)$ operations. To compute the first principal component costs ${\cal O}(n_{iter} M N)$, where $n_{iter}$ is the number of iterations:
\begin{eqnarray}\label{pow.it.1}
 &&\left[V^{(1)}_i\right]_{r+1} = {{\widetilde V_i} \over \sqrt{\sum_{j=1}^N {\widetilde V}_i^2}}\\
 &&{\widetilde V_i} = \sum_{j=1}^N \Psi_{ij}~\left[V^{(1)}_j\right]_{r} = {1\over M} \sum_{j=1}^N \sum_{s=1}^{M+1} Y_{is}~Y_{js}~\left[V^{(1)}_j\right]_{r}
 \label{power.it}
\end{eqnarray}
where $r$ labels the iterations.\footnote{\, Note that the sum over $s$ rums from 1 to $M+1$ in (\ref{power.it}).} Each iteration costs ${\cal O}(M N)$ operations. However, we need to compute $M$ principal components. This can be done as follows. Let
\begin{eqnarray}\label{Psi.Y}
 &&\Psi^{(a)}_{ij} = {1\over M} \sum_{s=1}^{M+1} Y_{is}^{(a)}~Y_{js}^{(a)}\\
 &&Y_{is}^{(a+1)} = Y_{is}^{(a)} - V^{(a)}_i \sum_{j=1}^N V^{(a)}_j ~ Y_{js}^{(a)}\\
 &&Y_{is}^{(1)} = Y_{is}
\end{eqnarray}
Note that $\Psi^{(1)}_{ij} = \Psi_{ij}$. The first principal component of $\Psi^{(a)}_{ij}$ -- which we can compute using the power iterations method -- is the same as the $a$-th principal component $V^{(a)}_i$ of $\Psi_{ij}$. Each such computation costs ${\cal O}(n^{(a)}_{iter}MN)$ operations. So, computing the first $M$ principal components costs ${\cal O}(n^{tot}_{iter}MN)$ operations, where $n^{tot}_{iter} = \sum_{a=1}^M n^{(a)}_{iter}$, and typically $n^{tot}_{iter}\gg M$. Source code for this procedure is given in Appendix \ref{app.B}. Table \ref{table.prin.comp} gives an analysis for a time series with $M = 19$ for $N = 2,339$ stock returns (close-to-close). The results show that $n^{tot}_{iter}\gg M$ for a reasonable computational precision, so the cost of computing the $M$ principal components is substantially greater than ${\cal O}(M^2 N)$ operations. The power iterations method is not cheap...\footnote{\, Also see, e.g., \cite{GB} and references therein.}

{}The above procedure simply amounts to successively removing the already-computed principal components from $\Psi_{ij}$. Indeed, $\Psi^{(a+1)}_{ij} = \Psi^{(a)}_{ij} - \lambda^{(a)} V^{(a)}_i V^{(a)}_j$. The reason to express $\Psi^{(a)}_{ij}$ via $Y^{(a)}_{is}$ in (\ref{Psi.Y}) is so we have a factorized form, which leads to a much smaller number of operations required to multiply this matrix by the iteration $\left[V^{(a)}_j\right]_{r}$. Note that $\Psi^{(M)}_{ij} = \lambda^{(M)}V^{(M)}_iV^{(M)}_j$, so irrespective of $\left[V^{(M)}_i\right]_{init}$, we get $V^{(M)}_i$ via (\ref{pow.it.1}) and (\ref{power.it}) (with $V^{(1)}_i$, $\Psi_{ij}$ and $Y_{is}$ replaced by $V^{(M)}_i$, $\Psi^{(M)}_{ij}$ and $Y^{(M)}_{is}$) right at the first iteration (hence only 2 iterations in the last row in Table \ref{table.prin.comp}).

\subsection{Computing Principal Components without Iterations}\label{sub.no.iter}

{}However, when $M \ll N$, we can compute the $M$ principal components of the sample correlation matrix without any costly iterations (involving $N$-vectors) and the cost is ${\cal O}(M^2 N)$ operations. We start with (\ref{CorMat}). Let (in matrix notation) $\phi = \varphi~\varphi^T$, where $\varphi$ is the Cholesky decomposition of $\phi$. ($\varphi_{ss} = \sqrt{(s+1)/sM}$; $\varphi_{ss^\prime} = 1/\sqrt{s^\prime(s^\prime+1)M}$ for $s > s^\prime$; $\varphi_{ss^\prime}=0$ for $s < s^\prime$.) Let ${\widetilde Y} = Y\varphi$. Then we have
\begin{equation}
 \Psi_{ij} = \sum_{s=1}^M {\widetilde Y}_{is}~{\widetilde Y}_{js}
\end{equation}
The columns of ${\widetilde Y}_{is}$ are not orthonormal. Let
\begin{equation}
 G_{ss^\prime} = \sum_{i=1}^N {\widetilde Y}_{is}~{\widetilde Y}_{is^\prime}
\end{equation}
We can readily find its eigenpairs. (This costs only ${\cal O}(M^3)$ operations.\footnote{\, Technically, this involves iterations, but no costly iterations involving $N$-vectors ($M\ll N$).}) Let the eigenvalues be $\rho^{(a)}$ and the principal components be $U^{(a)}_s$, $a=1,\dots,M$. Then the first $M$ principal components of $\Psi_{ij}$ are given by (this costs ${\cal O}(M^2 N)$ operations):
\begin{equation}\label{V.Y}
 V^{(a)}_i = {1\over \sqrt{\rho^{(a)}}}\sum_{s = 1}^M {\widetilde Y}_{is}~U^{(a)}_s
\end{equation}
Indeed, we have (in matrix notation) $\Psi~V^{(a)} = \lambda^{(a)}~V^{(a)}$, where $\lambda^{(a)} = \rho^{(a)}$, and $\sum_{i=1}^N V^{(a)}_i~V^{(b)}_i = \delta_{ab}$. The R source code for this method is given in Appendix \ref{app.C}.

\subsection{Statistical Risk Model = Deformation}

{}The purpose of the last subsection is not only to discuss an efficient method for computing eigenpairs, but also to rewrite the statistical risk model given by (\ref{PC}) and (\ref{PC.xi}) directly in terms of the (normalized demeaned) returns $Y_{is}$. Thus, using (\ref{V.Y}) we have\footnote{\, Note that this holds for any $M\leq N$, not just $M\ll N$.}
\begin{eqnarray}\label{PC.def}
 &&\Gamma_{ij} = \xi_i^2~\delta_{ij} + \sum_{s,s^\prime = 1}^M Y_{is}~{\widetilde\phi}_{ss^\prime}~Y_{js^\prime}\\
 &&\xi_i^2 = 1 - \sum_{s,s^\prime = 1}^M Y_{is}~{\widetilde\phi}_{ss^\prime}~Y_{is^\prime}
\end{eqnarray}
where
\begin{eqnarray}\label{phi.def}
 &&{\widetilde\phi}_{ss^\prime} = \sum_{r,r^\prime=1}^M \varphi_{sr}~F_{rr^\prime}~\varphi_{s^\prime r^\prime}\\
 &&F_{rr^\prime} = \sum_{a=1}^K U^{(a)}_r~U^{(a)}_{r^\prime}
\end{eqnarray}
When $K = M$, we have $F_{rr^\prime}=\delta_{rr^\prime}$ and ${\widetilde\phi}_{ss^\prime}={\phi}_{ss^\prime}$, so $\xi_i^2\equiv 0$ and $\Gamma_{ij} = \Psi_{ij}$. So, the statistical risk model (\ref{PC}), as it can be rewritten via (\ref{PC.def}), is nothing but a {\em deformation} (or regularization) of the sample correlation matrix $\Psi_{ij}$ given by (\ref{CorMat}): we deform $\phi_{ss^\prime}\rightarrow{\widetilde\phi}_{ss^\prime}$ thereby reducing the factor risk contribution into the total risk and replace the deficit by the specific risk. In this regard, note that we can consider more general deformations of the sample correlation matrix of the form
\begin{equation}\label{Psi.twiddle}
 {\widetilde \Psi}_{ij} = \Delta_{ij} + \sum_{s,s^\prime = 1}^M Y_{is}~{\widetilde\phi}_{ss^\prime}~Y_{js^\prime}
\end{equation}
subject to the requirements that ${\widetilde \Psi}_{ij}$ be positive-definite and ${\widetilde \Psi}_{ii}\equiv 1$. However, in practice, sticking to our basic premise that there is no information available beyond the returns (i.e., no style or industry factors can be constructed), there is no choice but to take diagonal $\Delta_{ij}$, which is then completely fixed. So, we are left with the choice of deforming $\phi_{ss^\prime}\rightarrow{\widetilde\phi}_{ss^\prime}$. And (\ref{phi.def}) is just {\em one} of myriad such deformations.

{}In fact, (\ref{phi.def}) is not even the simplest such deformation. It is the one that arises in {\em traditional} statistical risk models based on principal components. However, a choice to work with the principal components is by large simply a matter of taste (or even habit). It is just one basis, which a priori is not necessarily better or worse than any other basis. The {\em intuitive} justification behind a statistical risk model of the form (\ref{PC}) is clear: we keep the first $K$ principal components with the largest contributions to the sample correlation matrix -- based on the fact that the eigenvalues $\lambda^{(1)} > \lambda^{(2)} > \dots$ -- and replace the deficit (on the diagonal) by the specific risk. On the surface it all appears to make sense. However, the principal components beyond the first one are not stable out-of-sample, and this instability is inherited from that of the off-diagonal elements of the sample correlation matrix (i.e., pair-wise correlations). The first principal component also depends on the latter; however, for large $N$, in the leading approximation we have $V^{(1)}_i\approx 1/\sqrt{N}$ (the so-called ``market mode" -- see, e.g., \cite{CFM} and references therein), which is by definition stable, albeit subleading corrections are not. In any event, there is no reason to limit ourselves to the deformations of the form (\ref{phi.def}).

\subsection{``Nontraditional" Statistical Risk Models}

{}While a priori we can consider an arbitrary deformation $\phi_{ss}\rightarrow{\widetilde \phi}_{ss^\prime}$ (subject to the requirement that $\xi_i^2 > 0$), in practice it ought to be reasonable in the sense that it has to work out-of-sample. In this regard, keeping the first $K$ principal component of $\Psi_{ij}$ can be argued to be reasonable in the sense that, while the principal components themselves are not stable out-of-sample (except for the quasi-stable first principal component -- see above), tossing the higher principal components makes sense as their contributions are suppressed by the corresponding eigenvalues. This yields the deformation (\ref{phi.def}), which involves the matrix $F_{ss^\prime}$ constructed from the returns $Y_{is}$.

{}So, can we deform $\phi_{ss^\prime}$ directly, without any reference to the returns $Y_{is}$? The simplest such deformation is
\begin{eqnarray}\label{def.sh}
 &&{\widetilde\phi}_{ss^\prime} = \left(1 - q\right) \phi_{ss^\prime}\\
 &&\xi_i^2 \equiv q\label{xi.sh}
\end{eqnarray}
This is nothing but shrinkage \cite{LW}. Here we are shrinking the sample correlation matrix (as opposed to the sample covariance matrix, which is what is usually done); $0 < q < 1$ is the ``shrinkage constant"; the ``shrinkage target" is the diagonal $N\times N$ unit matrix. So, in this case we have $K = M$, the factor loadings matrix is given by (\ref{FLM.PC}), but instead of (\ref{CorMat.PC}) we have the factor covariance matrix $\Phi_{AB} = \left(1-q\right)\delta_{AB}$. So, shrinkage is a factor model \cite{shrunk}.

{}Unlike in the traditional statistical risk models, where we need to fix the number of factors $K$, in shrinkage the number of factors is fixed ($K=M$ for the diagonal shrinkage target; $\Delta_{ij} = \xi_i^2\delta_{ij}$ -- see below), but we must fix the shrinkage constant $q$ instead.\footnote{\, A way to fix $q$ is discussed in \cite{LW}.} However, contrary to an apparent common misconception, the value of $q$ makes little difference if the number of returns $N\gg 1$. This is because in this case optimization using statistical models is well-approximated by a weighted regression.

\subsection{Optimization $\approx$ Regression}\label{sub.shrinkage}

{}Optimization involves the inverse ${\widehat \Gamma}^{-1}_{ij}$ of the model covariance matrix ${\widehat\Gamma}_{ij} = \sigma_i\sigma_j\Gamma_{ij}$, where $\Gamma_{ij}$ is given by (\ref{fac.mod}).  For our purposes here it is convenient to rewrite $\Gamma_{ij}$ via $\Gamma_{ij} = \xi_i\xi_j\gamma_{ij}$, where
\begin{equation}
 \gamma_{ij} = \delta_{ij} + \sum_{A=1}^K \beta_{iA}~\beta_{jA}
\end{equation}
and $\beta_{iA} = {\widetilde \beta}_{iA} / \xi_i$. Here (in matrix notation) ${\widetilde\beta} = \Omega~{\widetilde \Phi}$, and ${\widetilde\Phi}$ is the Cholesky decomposition of $\Phi$, so ${\widetilde\Phi}~{\widetilde \Phi}^T = \Phi$. So, we have $\Gamma^{-1}_{ij} = \gamma^{-1}_{ij}/\xi_i\xi_j$, where
\begin{eqnarray}
 &&\gamma^{-1}_{ij} = \delta_{ij} - \sum_{A,B=1}^K \beta_{iA}~Q^{-1}_{AB}~\beta_{jB}\\
 &&Q_{AB} = \delta_{AB} + q_{AB}\\
 &&q_{AB} = \sum_{i=1}^N \beta_{iA}~\beta_{iB}
\end{eqnarray}
It then follows that, if all $q_{AA} = \sum_{i=1}^N \beta_{iA}^2 \gg 1$, then $Q_{AB}\approx q_{AB}$ and
\begin{eqnarray}
 &&\gamma^{-1}_{ij}\approx\delta_{ij} - \sum_{A,B=1}^K \beta_{iA}~q^{-1}_{AB}~\beta_{jB} =  \delta_{ij} - {1\over\xi_i\xi_j}\sum_{A,B=1}^K \Omega_{iA}~{\widetilde q}^{-1}_{AB}~\Omega_{jB}\\
 &&{\widetilde q}_{AB} = \sum_{i = 1}^N {1\over\xi^2_i}~\Omega_{iA}~\Omega_{iB}
\end{eqnarray}
I.e., in this case $\gamma^{-1}_{ij}$ is (approximately) independent of the factor covariance matrix. Furthermore, for an arbitrary vector $Z_i$, we have
\begin{equation}\label{sh.reg}
 \sum_{i=1}^N{\widehat \Gamma}^{-1}_{ij}~Z_j \approx  \omega_i\left[Z_i - \sum_{j = 1}^N \sum_{A,B = 1}^K {\widehat \Omega}_{iA}~{\widetilde q}^{-1}_{AB}~{\widehat \Omega}_{jB}~\omega_j~Z_j\right] = \omega_i~\varepsilon_i
\end{equation}
Here: $\omega_i = 1/\sigma_i^2\xi_i^2$; ${\widehat \Omega}_{iA} = \sigma_i\Omega_{iA}$; and $\varepsilon_i$ are the residuals of the cross-sectional weighted regression (without the intercept) of $Z_i$ over ${\widehat \Omega}_{iA}$ with the weights $\omega_i$. So, in this regime, the optimization based on shrinkage reduces to a weighted regression.

{}The question is why -- or, more precisely, when -- all $q_{AA}\gg 1$. This is the case when: i) $N$ is large, and ii) there is no ``clustering" in the vectors $\beta_{iA}$. That is, we do not have vanishing or small values of $\beta_{iA}^2$ for most values of the index $i$ with only a small subset thereof having $\beta_{iA}^2\gsim 1$. Without ``clustering", to have $q_{AA}\lsim 1$, we would have to have $\beta^2_{iA}\ll 1$, i.e., $\gamma_{ij}$ and consequently $\Gamma_{ij}$ would be almost diagonal. And such ``clustering" is certainly absent if $\Omega_{iA}$ are the $M$ principal components of the sample correlation matrix, so the above approximation holds in the case of shrinkage. The matrix ${\widehat \Omega}_{iA} = V_i^{(A)}$, $A=1,\dots,M$, and is independent of the shrinkage constant $q$. The regression weights $\omega_i = 1/q\sigma_i^2$ do depend on $q$, but this dependence does not affect the desired holdings $H_i$ in (\ref{H.opt}) as $q$ simply rescales the overall normalization coefficient $\eta$ in (\ref{H.opt}). That is, for large $N$, the desired holdings based on shrinkage are approximately independent of the shrinkage constant $q$. Table \ref{table.shrinkage} gives the simulation results\footnote{\, With the bounds (\ref{liq}) -- not including the bounds does not change the qualitative picture.} for various values of $q$. For $q=0$ we have $\Gamma_{ij} = \Psi_{ij}$, which is singular, so in this case the computation is done via the weighted regression -- see (\ref{sh.reg}).\footnote{\, With no bounds the desired holdings (\ref{H.opt}) can be obtained in two steps: first we regress the expected returns $E_i$ over ${\widehat \Omega}_{iA}$ (without the intercept) and weights $\omega_i = 1/\sigma_i^2$, and then we demean the residuals. With the bounds these two steps cannot be separated. So, we have two options. We can simply set $q$ to a small number (e.g., $q = 10^{-6}$) and use the R code in Appendix C of \cite{Het}. Or we can modify said code by (straightforwardly) replacing the optimization procedure therein via a weighted regression thereby arriving at ``bounded regression with linear constraints". The result is slightly better if we simply regress $E_i$ over ${\widehat \Omega}_{iA}$ {\em with} the intercept and weights $\omega_i = 1/\sigma_i^2$, in which case we have ROC 40.74\%, SR 13.86, CPS 1.81.\label{fn.add.int}} Not only is the value of the shrinkage constant immaterial, but the shrinkage based models sizably underperform the traditional statistical risk model with $K$ fixed via the ``minimization" algorithm of \cite{Het}.

{}To be clear, let us note that our observation that for large $N$ the optimization reduces to the weighted regression also applies to the traditional statistical risk models. The difference between the latter and shrinkage is that i) fewer than $M$ principal components are used as the columns of the $N\times K$ factor loadings matrix $\Omega_{iA}$ (i.e., $K < M$), and ii) the specific variances $\xi_i^2$ are no longer uniform (cf. (\ref{xi.sh})).

{}Let us mention that the shrinkage deformation (\ref{def.sh}) can be straightforwardly generalized via\footnote{\, This choice uniquely preserves the ${\bf Z}_M$ permutational symmetry under which $s \rightarrow s+1$, $s^\prime \rightarrow s^\prime+1$, and $s\,(s^\prime)>M$ is identified with $s-M$ ($s^\prime - M$).} ${\widetilde\phi}_{ss^\prime} = {1\over M}\left(\alpha\delta_{ij} + \beta u_s u_{s^\prime}\right)$ ($u_s\equiv 1$). In this case we still have the same $M$ risk factors,\footnote{\, I.e., the first $M$ principal components of $\Psi_{ij}$ -- the deformation ${\phi}_{ss^\prime} \rightarrow {\widetilde \phi}_{ss^\prime}$ simply rotates the basis so long as ${\widetilde \phi}_{ss^\prime}$ is nonsingular.} but we no longer have uniform $\xi_i^2 = 1-\alpha-{{\beta-\alpha}\over M}\left[\sum_{s=1}^M Y_{is}\right]^2$. For $\beta = \alpha = 1 - q$ this gives the standard shrinkage. Generally, we can have $\beta\neq \alpha$.

{}Finally, we can actually increase the number of risk factors beyond $M$. This can be done by considering $\Delta_{ij}$ in (\ref{Psi.twiddle}) that itself is a factor model. As before, let us continue assuming that we cannot construct any nontrivial style factors and there is no industry classification either. We can still construct a 1-factor model for $\Delta_{ij}$ with the intercept as the factor. If we choose the deformation ${\widetilde\phi}_{ss^\prime}=(1-q){\phi}_{ss^\prime}$, then we can set $\Delta_{ij} = \xi_i^2~\delta_{ij} + q~\rho~\nu_i~\nu_j$ ($\nu_i\equiv 1$, $|\rho| < 1$), so the factor model covariance matrix now reads\footnote{\, This is the shrinkage model of \cite{LW} where the shrinkage target corresponds to uniform correlations. If we take ${\widetilde\phi}_{ss^\prime} = {1\over M}\left(\alpha\delta_{ij} + \beta u_s u_{s^\prime}\right)$, then $\xi_i^2$ are nonuniform for $\beta\neq\alpha$.}
\begin{equation}
 \Gamma_{ij} = \xi_i^2~\delta_{ij} + q~\rho~\nu_i~\nu_j + (1-q) \sum_{s,s^\prime = 1}^M Y_{is}~{\phi}_{ss^\prime}~Y_{js^\prime}
\end{equation}
where $\xi_i^2 \equiv q(1-\rho)$. So, we have $M+1$ factors, the $M$ principal components plus the intercept. For large $N$, as above, the optimization reduces to a weighted regression, except that now it is with the intercept. The result is essentially independent of the values of $q$ and $\rho$ (see Table \ref{table.shrink.rho}) and expectedly somewhat better than in Table \ref{table.shrinkage}.

\section{Concluding Remarks}\label{sec.conc}

{}To begin with, let us tie a loose end. We discussed the algorithms for fixing the number of statistical factors $K$ (the ``minimization" and eRank based algorithms and their variations). Here we can ask: how do we know that, say, the ``minimization" based algorithm works better than picking some fixed value of $K$? This is tricky.

{}Indeed, how do we pick such ``optimal" fixed $K$? We can do this by simply running $M-1$ backtests with fixed $K=1,\dots,M-1$ for a given alpha and picking the value of $K$ that performs best. However, this ``optimal" value would be in-sample. There is no guarantee that it will work out-of-sample. Furthermore, generally it will vary from alpha to alpha. The aforementioned ``minimization" and eRank based algorithms by construction are oblivious to a choice of a sample, and even if they do not necessarily produce the ``optimal" value of $K$ for any given sample, they work for any sample. So, here we can ask whether they produce reasonable results for a given sample. Tables \ref{table.fixed.pc} and \ref{table.fixed.pc.bounds} give the simulation results for various fixed values of $K$ without and with the bounds (\ref{liq}), respectively. Looking at these results (especially those with the bounds, as any outperformance without the bounds should be taken with a grain of salt) it is evident that the fixed $K$ performance peaks around $K\approx 10$, while the ``minimization" based algorithm compares closer to $K$ between 12 and 13. This is consistent with the first row of Table \ref{table3}. The important thing is that the ``minimization" based algorithm produces results that are close to the in-sample ``optimal" results for fixed $K=10$.

{}In this regard it is instructive to run the following two series of backtests: i) taking the maximum $K_1=M$ risk factors fixed but ad hoc basing the specific risks on the $K$-factor model with varying $K$ (see Table \ref{table.20.pc.xi.bounds} and Figure 1); and ii) varying the number $K$ of the risk factors but ad hoc setting the specific risk equal the in-sample risk, i.e., $\xi_i^2 \equiv 1$ (see Table \ref{table.pc.tv.bounds} and Figure 2). The simulation results indicate that both the specific risk and the number of risk factors make a difference. The performance in Table \ref{table.20.pc.xi.bounds} peaks
around $K=13$ (also see Figure 1), which is the (approximate) number of risk factors fixed via the ``minimization" based algorithm of \cite{Het}. Not surprisingly, the Sharpe ratio in Table \ref{table.pc.tv.bounds} (also see Figure 2) improves as $K$ increases. However, the improvement rate slows down at higher $K$, so the specific risk effect is dominant. The performance in both Table \ref{table.20.pc.xi.bounds} and Table \ref{table.pc.tv.bounds} is worse than for the ``minimization" based algorithm (see Table \ref{table1}).

{}We already mentioned above (see footnote \ref{fn.xi.distrib}) that the ``minimization" based algorithm produces $K$ somewhat higher than $M/2$ due to a typically skewed $\xi_i^2$ distribution for a typical universe of stocks we dealt with in our backtests. Based on the above discussion, it might be tempting to simply set $K$ to the floor or cap of (or just rounded) $M/2$. Again, such a heuristic might work in-sample for some alphas, but not generally. Thus, if the underlying returns are highly correlated, a typical value of $K$ produced by the ``minimization" and eRank algorithms can be substantially lower than $M/2$ (including $K=1$). In such cases we can use the variation described in Subsection \ref{sub.k.prime}, and then a priori there is no reason to expect $K \approx M/2$. A safer path would appear to be to use different methods, see if they produce consistent results out-of-sample, and pick one based thereon.

{}Let us also mention that in the R code in both functions in Appendix \ref{app.A} we use the built-in R function {\tt{\small eigen()}} to compute eigenpairs. For large $N$ it is more efficient to replace it by the {\tt{\small qrm.calc.eignen.eff()}} function given in Appendix \ref{app.C}.

{}From our discussion above and backtests it is evident that ``nontraditional" statistical risk models such as shrinkage underperform the traditional statistical risk models. The reason why is that in ``nontraditional" models the rank of ${\widetilde\phi}_{ss^\prime}$ equals $M$, while in traditional models it is reduced, which yields nonuniform specific risks.

{}Furthermore, in the case of equity portfolios for which well-built and granular enough industry classifications are available, statistical risk models simply have no chance against risk models utilizing an industry classification such as heterotic risk models \cite{Het} or heterotic CAPM \cite{HetPlus}. The reason for this is twofold: i) industry factors are much more ubiquitous (thereby covering much more of the relevant risk space); and ii) principal components beyond the first one are unstable out-of-sample. In contrast, heterotic risk models use much more stable first principal components within each ``cluster" (e.g., BICS sub-industry), while heterotic CAPM uses a style factor.

\appendix

\section{R Code for Statistical Risk Models}\label{app.A}

{}In this appendix we give the R (R Package for Statistical Computing, http://www.r-project.org) source code for building a purely statistical risk model (principal components) based on the algorithm we discuss in Sections \ref{sec.2} and \ref{sub.fix.K}, including the ``minimization" and eRank based algorithms for fixing the number of factors $K$ in Section \ref{sub.fix.K}. The two functions below are essentially self-explanatory and straightforward.

{}The function {\tt{\small qrm.cov.pc(ret, use.cor = T, excl.first = F)}} corresponds to the ``minimization" based method for fixing $K$. The input is: i) {\tt{\small ret}}, an $N\times d$ matrix of returns (e.g., daily close-to-close returns), where $N$ is the number of returns, $d = M+1$ is the number of observations in the time series (e.g., the number of trading days), and the ordering of the dates is immaterial; ii) {\tt{\small use.cor}}, where for {\tt{\small TRUE}} (default) the risk factors are computed based on the principal components of the sample correlation matrix $\Psi_{ij}$, whereas for {\tt{\small FALSE}} they are computed based on the sample covariance matrix $C_{ij}$; {\tt{\small excl.first}}, where for {\tt{\small TRUE}} the $K^\prime$ based method of Subsection \ref{sub.k.prime} is used. The output is a list: {\tt{\small result\$spec.risk}} is the specific risk $\xi_i$ (not the specific variance $\xi_i^2$) for {\tt{\small use.cor = F}}, and $\sigma_i\xi_i$ (where $\sigma_i = \sqrt{C_{ii}}$) for {\tt{\small use.cor = T}} (recall that in this case $\xi_i$ is the specific risk for the factor model for $\Psi_{ij}$, not $C_{ij}$); {\tt{\small result\$fac.load}} is the factor loadings matrix $\Omega_{iA}$ for {\tt{\small use.cor = F}}, and $\sigma_i\Omega_{iA}$ for {\tt{\small use.cor = T}} (recall that in this case $\Omega_{iA}$ is the factor loadings matrix for the factor model for $\Psi_{ij}$, not $C_{ij}$); {\tt{\small result\$fac.cov}} is the factor covariance matrix $\Phi_{AB}$ (with the normalization (\ref{FLM.PC}) for the factor loadings matrix, $\Phi_{AB} = \delta_{AB}$); {\tt{\small result\$cov.mat}} is the factor model covariance matrix $\Gamma_{ij}$  for {\tt{\small use.cor = F}}, and $\sigma_i\sigma_j{\Gamma}_{ij}$ for {\tt{\small use.cor = T}}; {\tt{\small result\$inv.cov}} is the matrix inverse to {\tt{\small result\$cov.mat}}; {\tt{\small result\$pc}} are the first $K$ principal components of a) $C_{ij}$ for {\tt{\small use.cor = F}}, and b) $\Psi_{ij}$ for {\tt{\small use.cor = T}}.

{}The second function is {\tt{\small qrm.erank.pc(ret, use.cor = T, do.trunc = F, k = 0, excl.first = F)}} and corresponds to the eRank based method for fixing $K$ for the default parameter {\tt{\small k = 0}}. The input is the same as in the {\tt{\small qrm.cov.pc()}} function except for the additional parameters {\tt{\small do.trunc = F}} and {\tt{\small k = 0}}. For a positive integer {\tt{\small k}} the code simply takes its value as the number of factors $K$. For {\tt{\small k = 0}} (default) the code uses the eRank method: if {\tt{\small do.trunc = F}} (default), then $K = \mbox{Round}(\mbox{eRank}(\cdot))$, while if {\tt{\small do.trunc = T}}, then $K = \mbox{floor}(\mbox{eRank}(\cdot))$. (The argument of $\mbox{eRank}(\cdot)$ is the matrix $C_{ij}$ if {\tt{\small use.cor = F}}, and the matrix $\Psi_{ij}$ if {\tt{\small use.cor = T}}). The output is the same as in the {\tt{\small qrm.cov.pc()}} function.\\
\\
{\tt{\small
\noindent qrm.cov.pc <- function (ret, use.cor = T, excl.first = F)\\
\{\\
\indent print("Running qrm.cov.pc()...")\\
\\
\indent tr <- apply(ret, 1, sd)\\
\indent if(use.cor)\\
\indent \indent ret <- ret / tr\\
\\
\indent d <- ncol(ret)\\
\indent x <- t(ret)\\
\indent x <- var(x, x)\\
\indent tv <- diag(x)\\
\indent x <- eigen(x)\\
\indent if(excl.first)\\
\indent \{\\
\indent \indent k1 <- 2\\
\indent \indent y1 <- sqrt(x\$values[1]) * matrix(x\$vectors[, 1], nrow(ret), 1)\\
\indent \indent x1 <- y1 \%*\% t(y1)\\
\indent \indent tv <- tv - diag(x1)\\
\indent \}\\
\indent else\\
\indent \{\\
\indent \indent k1 <- 1\\
\indent \indent x1 <- 0\\
\indent \}\\
\\
\indent g.prev <- 999\\
\indent for(k in k1:(d-1))\\
\indent \{\\
\indent \indent u <- x\$values[k1:k]\\
\indent \indent v <- x\$vectors[, k1:k]\\
\indent \indent v <- t(sqrt(u) * t(v))\\
\indent \indent x.f <- v \%*\% t(v)\\
\indent \indent x.s <- tv - diag(x.f)\\
\indent \indent z <- x.s / tv\\
\indent \indent g <- abs(sqrt(min(z)) + sqrt(max(z)) - 1)\\
\\
\indent \indent if(is.na(g))\\
\indent \indent \indent break\\
\\
\indent \indent if(g > g.prev)\\
\indent \indent \indent break\\
\\
\indent \indent g.prev <- g\\
\\
\indent \indent spec.risk <- sqrt(x.s)\\
\indent \indent if(excl.first)\\
\indent \indent \indent fac.load <- cbind(y1, v)\\
\indent \indent else\\
\indent \indent \indent fac.load <- v\\
\indent \indent fac.cov <- diag(1, k)\\
\indent \indent cov.mat <- diag(x.s) + x.f + x1\\
\indent \}\\
\\
\indent y.s <- 1 / spec.risk\^{}2\\
\indent v <- fac.load\\
\indent v1 <- y.s * v\\
\indent inv.cov <- diag(y.s) - v1 \%*\%\\
\indent \indent solve(diag(1, ncol(v)) + t(v) \%*\% v1) \%*\% t(v1)\\
\\
\indent if(use.cor)\\
\indent \{\\
\indent \indent spec.risk <- tr * spec.risk\\
\indent \indent fac.load <- tr * fac.load\\
\indent \indent cov.mat <- tr * t(tr * cov.mat)\\
\indent \indent inv.cov <- t(inv.cov / tr) / tr\\
\indent \}\\
\\
\indent result <- new.env()\\
\indent result\$spec.risk <- spec.risk\\
\indent result\$fac.load <- fac.load\\
\indent result\$fac.cov <- fac.cov\\
\indent result\$cov.mat <- cov.mat\\
\indent result\$inv.cov <- inv.cov\\
\indent result\$pc <- x\$vectors[, 1:ncol(fac.load)]\\
\indent result <- as.list(result)\\
\indent return(result)\\
\}\\
\\
\noindent qrm.erank.pc <- function (ret, use.cor = T, do.trunc = F,\\
\indent \indent k = 0, excl.first = F)\\
\{\\
\indent print("Running qrm.erank.pc()...")\\
\\
\indent calc.erank <- function(x, excl.first)\\
\indent \{\\
\indent \indent take <- x > 0\\
\indent \indent x <- x[take]\\
\indent \indent if(excl.first)\\
\indent \indent \indent x <- x[-1]\\
\indent \indent p <- x / sum(x)\\
\indent \indent h <- - sum(p * log(p))\\
\indent \indent er <- exp(h)\\
\indent \indent if(excl.first)\\
\indent \indent \indent er <- er + 1\\
\indent \indent return(er)\\
\indent \}\\
\\
\indent tr <- apply(ret, 1, sd)\\
\indent if(use.cor)\\
\indent \indent ret <- ret / tr\\
\\
\indent x <- t(ret)\\
\indent x <- var(x, x)\\
\indent tv <- diag(x)\\
\indent y <- eigen(x)\\
\indent if(k == 0)\\
\indent \{\\
\indent \indent er <- calc.erank(y\$values, excl.first)\\
\\
\indent\indent if(do.trunc)\\
\indent \indent \indent k <- trunc(er)\\
\indent \indent else\\
\indent \indent \indent k <- round(er)\\
\indent \}\\
\\
\indent k <- min(k, ncol(ret) - 2)\\
\indent fac.load <- t(t(y\$vectors[, 1:k]) * sqrt(y\$values[1:k]))\\
\indent fac.cov <- diag(1, k)\\
\indent x.f <- fac.load \%*\% t(fac.load)\\
\indent x.s <- tv - diag(x.f)\\
\indent spec.risk <- sqrt(x.s)\\
\indent cov.mat <- diag(x.s) + x.f\\
\\
\indent y.s <- 1 / spec.risk\^{}2\\
\indent v <- fac.load\\
\indent v1 <- y.s * v\\
\indent inv.cov <- diag(y.s) - v1 \%*\%\\
\indent \indent solve(diag(1, k) + t(v) \%*\% v1) \%*\% t(v1)\\
\\
\indent if(use.cor)\\
\indent \{\\
\indent \indent spec.risk <- tr * spec.risk\\
\indent \indent fac.load <- tr * fac.load\\
\indent \indent cov.mat <- tr * t(tr * cov.mat)\\
\indent \indent inv.cov <- t(inv.cov / tr) / tr\\
\indent \}\\
\\
\indent result <- new.env()\\
\indent result\$spec.risk <- spec.risk\\
\indent result\$fac.load <- fac.load\\
\indent result\$fac.cov <- fac.cov\\
\indent result\$cov.mat <- cov.mat\\
\indent result\$inv.cov <- inv.cov\\
\indent result\$pc <- y\$vectors[, 1:k]\\
\indent result <- as.list(result)\\
\indent return(result)\\
\}
}}

\section{R Code for Eigenpairs via Power Iterations}\label{app.B}

{}In this appendix we give the R source code for calculating the first $M$ eigenpairs of the sample correlation matrix $\Psi_{ij}$ based on the successive application of the power iterations method as in Section \ref{sub.pc}. The code below is essentially self-explanatory and straightforward as it simply follows the formulas in Section \ref{sub.pc}. It consists of a single function {\tt{\small qrm.calc.eigen(ret, k, prec = 1e-3)}}; {\tt{\small ret}} is an $N\times (M+1)$ matrix of returns; $N$ is the number of the underlying returns (e.g., alphas); $M+1$ is the number of data points in the time series (e.g., days); {\tt{\small k}} is the number of the desired first $k$ eigenpairs (in the decreasing order of the eigenvalues) to be computed (if $k > M$, only the first $M$ eigenpairs are computed); {\tt{\small prec}} is the convergence precision and is not the same as how close $\sum_{j=1}^N \Psi_{ij}V^{(a)}_j / \lambda^{(a)}V^{(a)}_i$ are to 1 or the approximate eigenvectors $V^{(a)}_i$ are to the true eigenvectors. That precision is lower owing to cumulative effects (due to sums, etc.). The output is a list: {\tt{\small result\$count}} is a $k$-vector whose elements are the numbers of iterations $n^{(a)}_{iter}$, $a=1,\dots,k$, for each eigenpair; {\tt{\small result\$value}} is a $k$-vector of the $k$ eigenvalues; and {\tt{\small result\$vector}} is an $N\times k$ matrix whose columns are the $k$ eigenvectors. The two lines {\tt{\small y[] <- rnorm(n, 0, 1)}} and {\tt{\small y <- y / sqrt(sum(y\^{}2))}} are optional and used here for the purpose of generating randomness in the runs in Table \ref{table.prin.comp}; more economical ways of setting the initial iterations for the eigenpairs with $a>1$ can be used.\\
\\
{\tt{\small
\noindent qrm.calc.eigen <- function(ret, k, prec = 1e-3)\\
\{\\
\indent pow.it.fac <- function(x, y, prec)\\
\indent \{\\
\indent \indent count <- 0\\
\indent \indent repeat\{\\
\indent \indent \indent count <- count + 1\\
\indent \indent \indent y.prev <- y\\
\indent \indent \indent y <- t(x) \%*\% y\\
\indent \indent \indent y <- x \%*\% y\\
\indent \indent \indent y <- y / sqrt(sum(y\^{}2))\\
\indent \indent \indent if(max(abs(y/y.prev - 1)) < prec)\\
\indent \indent \indent \indent break\\
\indent \indent \}\\
\indent \indent result <- new.env()\\
\indent \indent result\$count <- count\\
\indent \indent result\$value <- sum((t(x) \%*\% y)\^{}2)\\
\indent \indent result\$vector <- y\\
\indent \indent return(result)\\
\indent \}\\
\\
\indent n <- nrow(ret)\\
\indent m <- ncol(ret)\\
\indent k <- min(k, m - 1)\\
\indent cor.mat <- cor(t(ret), t(ret))\\
\indent x <- ret - rowMeans(ret)\\
\indent s <- sqrt(rowSums(x\^{}2))\\
\indent x <- x / s\\
\\
\indent count <- value <- rep(NA, k)\\
\indent vector <- matrix(NA, n, k)\\
\indent y <- matrix(1/sqrt(n), n, 1)\\
\\
\indent for(i in 1:k)\\
\indent \{\\
\indent \indent result <- pow.it.fac(x, y, prec)\\
\indent \indent count[i] <- result\$count\\
\indent \indent value[i] <- result\$value\\
\indent \indent vector[, i] <- v <- result\$vector\\
\\
\indent \indent for(j in 1:m)\\
\indent \indent \indent x[, j] <- x[, j] - v * sum(v * x[, j])\\
\\
\indent \indent y[] <- rnorm(n, 0, 1)\\
\indent \indent y <- y / sqrt(sum(y\^{}2))\\
\indent \}\\
\\
\indent result\$count <- count\\
\indent result\$value <- value\\
\indent result\$vector <- vector\\
\indent return(result)\\
\}
}}

\section{R Code for Eigenpairs without Iterations}\label{app.C}

{}In this appendix we give the R source code for calculating the first $M$ eigenpairs of the sample correlation matrix $\Psi_{ij}$ based on the no-iterations method discussed in Subsection \ref{sub.no.iter}. The code below is essentially self-explanatory and straightforward as it simply follows the formulas in Subsection \ref{sub.no.iter}. It consists of a single function {\tt{\small qrm.calc.eigen.eff(ret, calc.cor = F)}}; {\tt{\small ret}} is an $N\times (M+1)$ matrix of returns; $N$ is the number of the underlying returns; $M+1$ is the number of data points in the time series (e.g., days); for {\tt{\small calc.cor = F}} (default), the code computes the eigenpairs for the covariance matrix; for {\tt{\small calc.cor = T}} the code computes the eigenpairs for the correlation matrix. The method works only if $M\leq N$. The output is a list: {\tt{\small result\$values}} is an $M$-vector of the $M$ eigenvalues; and {\tt{\small result\$vectors}} is an $N\times M$ matrix whose columns are the $M$ eigenvectors.\\
\\
{\tt{\small
\noindent qrm.calc.eigen.eff <- function (ret, calc.cor = F)\\
\{\\
\indent calc.chol <- function(m)\\
\indent \{\\
\indent \indent x <- matrix(1, m, m)\\
\indent \indent x <- upper.tri(x, T)\\
\indent \indent z <- 1:m\\
\indent \indent x <- x + diag(z)\\
\indent \indent x <- x / sqrt(z * (z + 1))\\
\indent \indent x <- t(x) / sqrt(m)\\
\indent \indent return(x)\\
\indent \}\\
\\
\indent m <- ncol(ret) - 1\\
\indent n <- nrow(ret)\\
\\
\indent if(m > n)\\
\indent \indent stop("Too many observations...")\\
\\
\indent if(calc.cor)\\
\indent \indent ret <- ret / apply(ret, 1, sd)\\
\\
\indent y <- ret - rowMeans(ret)\\
\indent y <- y[, -(m + 1)]\\
\indent p <- calc.chol(m)\\
\indent y <- y \%*\% p\\
\indent q <- t(y) \%*\% y\\
\indent q <- eigen(q)\\
\indent q.vec <- q\$vectors\\
\indent q.val <- q\$values\\
\indent q.vec <- t(t(q.vec) / sqrt(q.val))\\
\indent y <- y \%*\% q.vec\\
\\
\indent result <- new.env()\\
\indent result\$values <- q.val\\
\indent result\$vectors <- y\\
\indent return(result)\\
\}
}}

\section{DISCLAIMERS}\label{app.D}

{}Wherever the context so requires, the masculine gender includes the feminine and/or neuter, and the singular form includes the plural and {\em vice versa}. The author of this paper (``Author") and his affiliates including without limitation Quantigic$^\circledR$ Solutions LLC (``Author's Affiliates" or ``his Affiliates") make no implied or express warranties or any other representations whatsoever, including without limitation implied warranties of merchantability and fitness for a particular purpose, in connection with or with regard to the content of this paper including without limitation any code or algorithms contained herein (``Content").

{}The reader may use the Content solely at his/her/its own risk and the reader shall have no claims whatsoever against the Author or his Affiliates and the Author and his Affiliates shall have no liability whatsoever to the reader or any third party whatsoever for any loss, expense, opportunity cost, damages or any other adverse effects whatsoever relating to or arising from the use of the Content by the reader including without any limitation whatsoever: any direct, indirect, incidental, special, consequential or any other damages incurred by the reader, however caused and under any theory of liability; any loss of profit (whether incurred directly or indirectly), any loss of goodwill or reputation, any loss of data suffered, cost of procurement of substitute goods or services, or any other tangible or intangible loss; any reliance placed by the reader on the completeness, accuracy or existence of the Content or any other effect of using the Content; and any and all other adversities or negative effects the reader might encounter in using the Content irrespective of whether the Author or his Affiliates is or are or should have been aware of such adversities or negative effects.

{}The R code included in Appendix \ref{app.A}, Appendix \ref{app.B} and Appendix \ref{app.C} hereof is part of the copyrighted R code of Quantigic$^\circledR$ Solutions LLC and is provided herein with the express permission of Quantigic$^\circledR$ Solutions LLC. The copyright owner retains all rights, title and interest in and to its copyrighted source code included in Appendix \ref{app.A}, Appendix \ref{app.B} and Appendix \ref{app.C} hereof and any and all copyrights therefor.



\newpage
\begin{figure}[ht]
\centerline{\epsfxsize 4.truein \epsfysize 4.truein\epsfbox{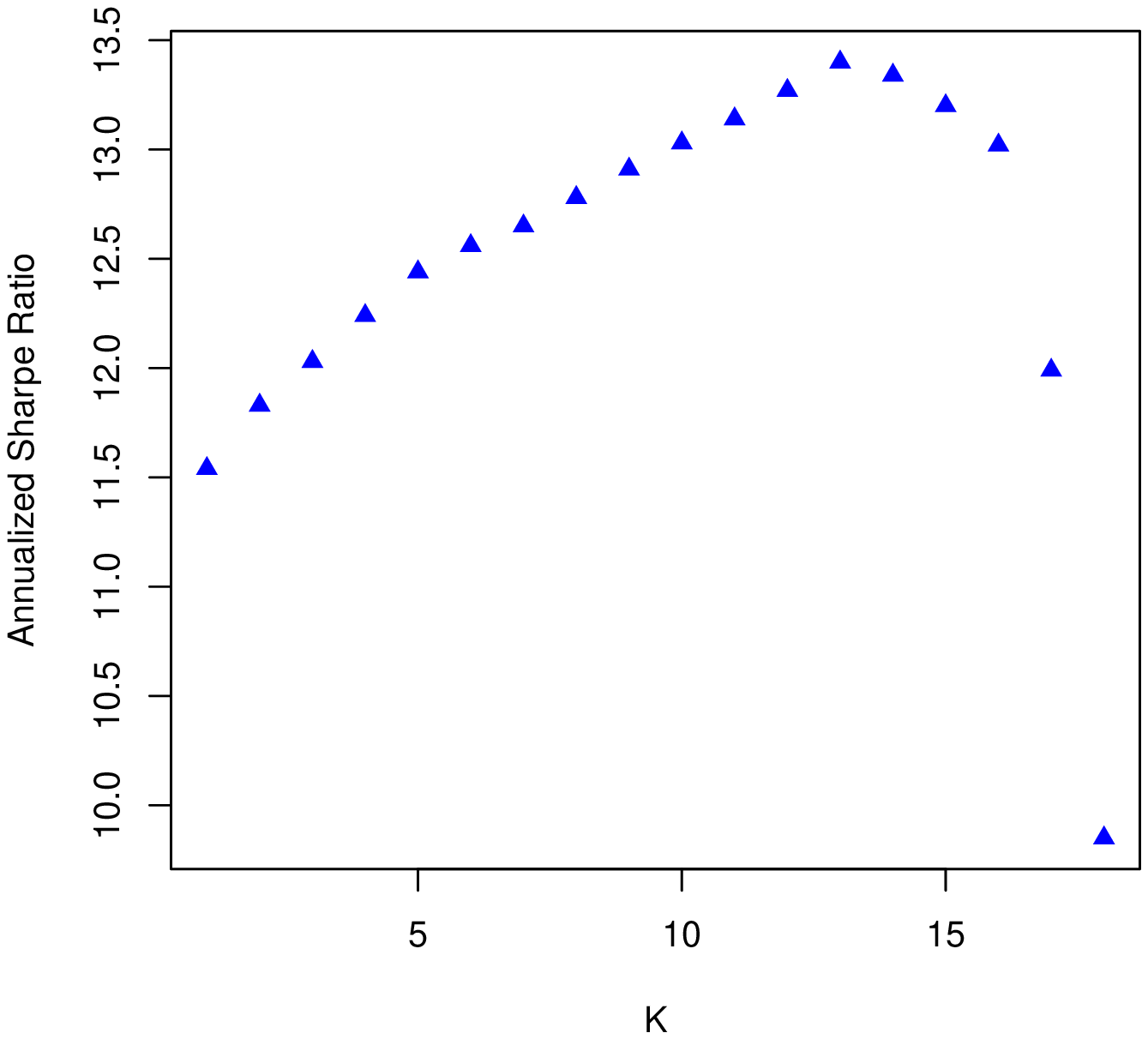}}
\noindent{\small {Figure 1. Graph of the values of the Sharpe ratio (SR) from Table \ref{table.20.pc.xi.bounds} vs. the number of risk factors $K$ (as defined in said table).}}
\end{figure}

\newpage
\begin{figure}[ht]
\centerline{\epsfxsize 4.truein \epsfysize 4.truein\epsfbox{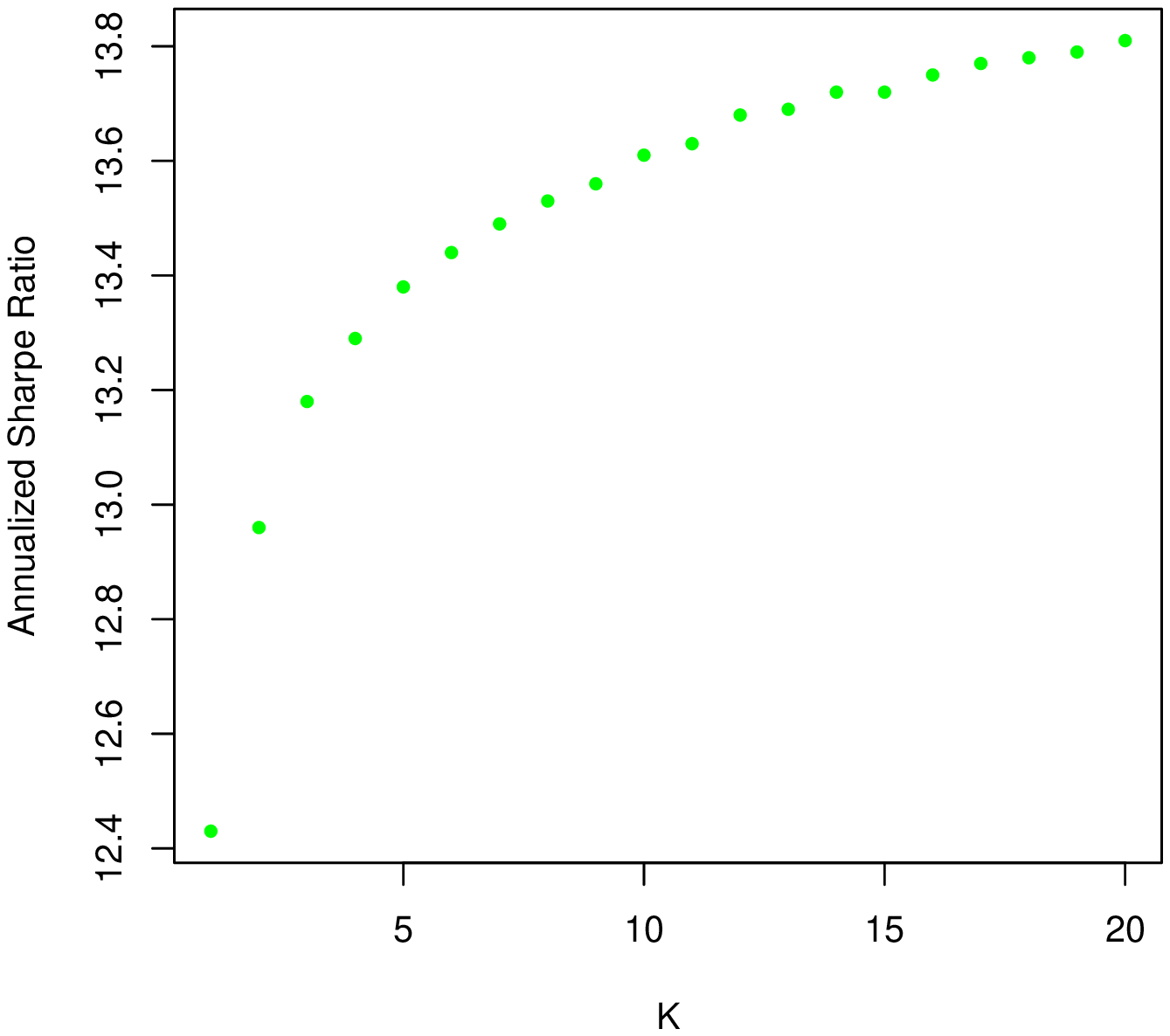}}
\noindent{\small {Figure 2. Graph of the values of the Sharpe ratio (SR) from Table \ref{table.pc.tv.bounds} vs. the number of risk factors $K$ (as defined in said table).}}
\end{figure}

\newpage
\begin{table}[ht]
\caption{Simulation results for the optimized alphas without bounds using the statistical risk models. See Subsection \ref{sub.backtests} for details. The result in the first row is the same as in \cite{Het}. We label the risk models M1-M6 for notational convenience in Table \ref{table3}.} 
\begin{tabular}{l l l l} 
\hline\hline 
Risk Model & ROC & SR & CPS\\[0.5ex] 
\hline 
M1: $K$, ``minimization" algorithm & 47.74\% & 11.88 & 2.26 \\
M2: $K = K^\prime + 1$, ``minimization" algorithm & 46.87\% & 11.59 & 2.23 \\
M3: $K = \mbox{Round}(\mbox{eRank}(\Psi))$ & 47.89\% & 11.20 & 2.28 \\
M4: $K = \mbox{floor}(\mbox{eRank}(\Psi))$ & 47.13\% & 11.10 & 2.24 \\
M5: $K = \mbox{Round}(\mbox{eRank}(\Psi^\prime)) + 1$ & 28.51\% & 4.67 & 1.32 \\
M6: $K = \mbox{floor}(\mbox{eRank}(\Psi^\prime)) + 1$ & 32.46\% & 5.53 & 1.49 \\[1ex] 
\hline 
\end{tabular}
\label{table1} 
\end{table}

\begin{table}[ht]
\caption{Simulation results for the optimized alphas with bounds. See Subsection \ref{sub.backtests} for details. The result in the first row is the same as
in \cite{Het}.} 
\begin{tabular}{l l l l} 
\hline\hline 
Risk Model & ROC & SR & CPS\\[0.5ex] 
\hline 
$K$, ``minimization" algorithm & 40.92\% & 14.33 & 1.96\\
$K = K^\prime + 1$, ``minimization" algorithm & 40.36\% & 14.04 & 1.94\\
$K = \mbox{Round}(\mbox{eRank}(\Psi))$ & 40.78\% & 14.01 & 1.96\\
$K = \mbox{floor}(\mbox{eRank}(\Psi))$ & 40.84\% & 14.06 & 1.96\\[1ex] 
\hline 
\end{tabular}
\label{table2} 
\end{table}

\begin{table}[ht]
\noindent
\caption{Summaries of various quantities from out backtests. The M1-M6 models are defined in Table \ref{table1}. 1st Qu. = 1st Quartile, 3rd Qu. = 3rd Quartile, StDev = standard deviation, MAD = mean absolute deviation. Summaries in the first 7 rows are computed based on the 60 data points corresponding to 60 21-trading-day intervals in our 1,260 trading-day backtesting interval (see Subsection \ref{sub.back}). The ``average correlation" ${\overline \Psi} = {1\over N^2}\sum_{i,j = 1}^N \Psi_{ij}$. The parameter $\zeta = \mbox{max}(|H_{is}| / 0.01~A_{is})$, where $\mbox{max}(\cdot)$ is cross-sectional for each date $s$ ($s$ takes 1,260 values) and the summary in the last row is taken over the 1,260 trading days.}
\begin{tabular}{l l l l l l l l l} 
\\
\hline\hline 
Quantity & Min & 1st Qu. & Median & Mean & 3rd Qu. & Max & StDev & MAD \\[0.5ex] 
\hline 
M1, $K$ & 6 & 11.75 & 12 & 12.17 & 13 & 14 & 1.57 & 1.48\\
M2, $K$ & 11 & 13 & 14 & 13.77 & 14 & 16 & 0.91 & 1.48\\
M3, $K$ & 3 & 10 & 12 & 11.93 & 14 & 17 & 3.31 & 2.97\\
M4, $K$ & 3 & 10 & 12 & 11.5 & 14 & 17 & 3.18 & 2.97\\
M5, $K$ & 17 & 19 & 19 & 18.92 & 19 & 19 & 0.33 & 0\\
M6, $K$ & 17 & 18 & 19 & 18.57 & 19 & 19 & 0.53 & 0\\
${\overline \Psi}$ (\%) & 11.68 & 22.86 & 31.14 & 32.93 & 38.43 & 74.78 & 13.44 & 11.46\\
M6, $\zeta$ & 3.26 & 16.66 & 25.22 & 29.66 & 37.39 & 189.5 & 19.19 & 14.35\\ [1ex] 
\hline 
\end{tabular}
\label{table3} 
\end{table}

\begin{table}[ht]
\noindent
\caption{The number of iterations $n^{(a)}_{iter}$, $a=1,\dots,M$, required to compute the first $M=19$ principal components of the sample correlation matrix based on a time series of $N = 2,339$ stock returns with $M + 1 = 20$ data points (trading days). There are 10 runs. In each run the initial iteration $[V^{(1)}_i]_{init}$ for the first principal component is taken as $[V^{(1)}_i]_{init} = u_i/\sqrt{N}$, where $u_i\equiv 1$ is the unit $N$-vector. For the other principal components the initial iterations are taken as random $N$-vectors with 0 mean and unit variance further normalized to have the quadratic norm 1, i.e., $\sum_{i=1}^N \left([V^{(a)}_i]_{init}\right)^2 = 1$, $a=2,\dots,M$. The mean (median) of the total iteration $n^{tot}_{iter}$ (i.e., of the last row) is 5,037 (5,038). See Subsection \ref{sub.pc} for details. We use the code in Appendix \ref{app.B} to generate this table. The convergence precision is set to the default, {\tt{\small prec = 1e-3}}. See Appendix \ref{app.B}.}
\begin{tabular}{l l l l l l l l l l l} 
\\
\hline\hline 
Run \#: & 1 & 2 & 3 & 4 & 5 & 6 & 7 & 8 & 9 & 10 \\[0.5ex] 
\hline 
$n^{(1)}_{iter}$ & 13 & 13 & 13 & 13 & 13 & 13 & 13 & 13 & 13 & 13\\
$n^{(2)}_{iter}$ & 24 & 24 & 25 & 24 & 23 & 24 & 25 & 23 & 23 & 22\\
$n^{(3)}_{iter}$ & 162 & 168 & 199 & 150 & 158 & 209 & 211 & 193 & 221 & 148\\
$n^{(4)}_{iter}$ & 181 & 165 & 167 & 211 & 219 & 158 & 160 & 207 & 185 & 166\\
$n^{(5)}_{iter}$ & 144 & 107 & 117 & 97 & 119 & 100 & 101 & 97 & 101 & 118\\
$n^{(6)}_{iter}$ & 457 & 298 & 371 & 287 & 387 & 387 & 392 & 359 & 426 & 448\\
$n^{(7)}_{iter}$ & 147 & 175 & 151 & 162 & 179 & 137 & 120 & 185 & 172 & 174\\
$n^{(8)}_{iter}$ & 430 & 309 & 372 & 443 & 361 & 440 & 437 & 366 & 395 & 412\\
$n^{(9)}_{iter}$ & 387 & 265 & 326 & 423 & 327 & 361 & 343 & 323 & 321 & 351\\
$n^{(10)}_{iter}$ & 222 & 209 & 217 & 209 & 273 & 229 & 316 & 208 & 210 & 234\\
$n^{(11)}_{iter}$ & 203 & 182 & 243 & 225 & 284 & 205 & 141 & 235 & 307 & 226\\
$n^{(12)}_{iter}$ & 261 & 242 & 235 & 288 & 365 & 296 & 286 & 284 & 259 & 252\\
$n^{(13)}_{iter}$ & 456 & 607 & 489 & 711 & 598 & 726 & 520 & 620 & 552 & 645\\
$n^{(14)}_{iter}$ & 1059 & 1029 & 1122 & 1089 & 1092 & 1296 & 1190 & 1169 & 1032 & 988\\
$n^{(15)}_{iter}$ & 307 & 403 & 392 & 371 & 441 & 438 & 364 & 437 & 392 & 409\\
$n^{(16)}_{iter}$ & 126 & 144 & 153 & 127 & 161 & 147 & 155 & 132 & 197 & 150\\
$n^{(17)}_{iter}$ & 156 & 192 & 203 & 157 & 206 & 177 & 167 & 171 & 197 & 177\\
$n^{(18)}_{iter}$ & 81 & 80 & 87 & 67 & 82 & 99 & 77 & 80 & 114 & 83\\
$n^{(19)}_{iter}$ & 2 & 2 & 2 & 2 & 2 & 2 & 2 & 2 & 2 & 2\\
$n^{tot}_{iter}$ & 4818 & 4614 & 4884 & 5056 & 5290 & 5444 & 5020 & 5104 & 5119 & 5018\\ [1ex] 
\hline 
\end{tabular}
\label{table.prin.comp} 
\end{table}

\begin{table}[ht]
\caption{Simulation results for the optimized alphas (with bounds) based on shrinkage with a diagonal shrinkage target for various values of the shrinkage constant $q$. See Subsection \ref{sub.shrinkage} for details.} 
\begin{tabular}{l l l l} 
\hline\hline 
$q$ & ROC & SR & CPS\\[0.5ex] 
\hline 
$10^{-6}$ & 40.67\% & 13.84 & 1.81\\
0.3 & 40.66\% & 13.83 & 1.81\\
0.6 & 40.66\% & 13.82 & 1.81\\
0.9 & 40.65\% & 13.81 & 1.81\\[1ex] 
\hline 
\end{tabular}
\label{table.shrinkage} 
\end{table}

\begin{table}[ht]
\caption{Simulation results for the optimized alphas (with bounds) based on shrinkage with a uniform correlation shrinkage target for various values of the shrinkage constant $q$ and the correlation $\rho$. See Subsection \ref{sub.shrinkage} for details.} 
\begin{tabular}{l l l l l} 
\hline\hline 
$q$ & $\rho$ & ROC & SR & CPS\\[0.5ex] 
\hline 
$10^{-6}$ & 0.1 & 41.03\% & 14.10 & 1.83\\
0.3 & 0.1 & 41.02\% & 14.09 & 1.83\\
0.6 & 0.1 & 41.01\% & 14.07 & 1.82\\
0.9 & 0.1 & 40.92\% & 13.92 & 1.82\\
$10^{-6}$ & 0.5 & 41.03\% & 14.11 & 1.83\\
0.3 & 0.5 & 41.03\% & 14.10 & 1.82\\
0.6 & 0.5 & 41.02\% & 14.09 & 1.82\\
0.9 & 0.5 & 40.98\% & 14.01 & 1.82\\
$10^{-6}$ & 0.9 & 41.03\% & 14.11 & 1.82\\
0.3 & 0.9 & 41.03\% & 14.11 & 1.82\\
0.6 & 0.9 & 41.03\% & 14.11 & 1.82\\
0.9 & 0.9 & 41.02\% & 14.09 & 1.82\\[1ex] 
\hline 
\end{tabular}
\label{table.shrink.rho} 
\end{table}

\begin{table}[ht]
\caption{Simulation results for the optimized alphas without bounds for fixed values of the number of risk factors $K$. See Section \ref{sec.conc} for details.} 
\begin{tabular}{l l l l} 
\hline\hline 
$K$ & ROC & SR & CPS\\[0.5ex] 
\hline 
1 & 46.69\% & 11.17 & 2.18\\
2 & 47.77\% & 11.69 & 2.24\\
3 & 47.83\% & 11.90 & 2.25\\
4 & 47.83\% & 11.90 & 2.26\\
5 & 48.26\% & 12.15 & 2.28\\
6 & 48.48\% & 11.89 & 2.29\\
7 & 48.73\% & 11.93 & 2.31\\
8 & 48.71\% & 12.02 & 2.31\\
9 & 48.68\% & 12.01 & 2.30\\
10 & 48.66\% & 12.16 & 2.30\\
11 & 48.21\% & 11.88 & 2.28\\
12 & 48.22\% & 11.79 & 2.29\\
13 & 47.54\% & 11.66 & 2.26\\
14 & 46.47\% & 10.98 & 2.21\\
15 & 45.90\% & 10.51 & 2.18\\
16 & 45.19\% & 10.56 & 2.15\\
17 & 45.96\% & 10.04 & 2.19\\
18 & 40.57\% & 6.98 & 1.89\\
19 & 27.78\% & 4.48 & 1.29\\[1ex] 
\hline 
\end{tabular}
\label{table.fixed.pc} 
\end{table}

\begin{table}[ht]
\caption{Simulation results for the optimized alphas with bounds for fixed values of the number of risk factors $K$. See Section \ref{sec.conc} for details.} 
\begin{tabular}{l l l l} 
\hline\hline 
$K$ & ROC & SR & CPS\\[0.5ex] 
\hline 
1 & 39.14\% & 12.97 & 1.85\\
2 & 40.20\% & 13.64 & 1.90\\
3 & 40.64\% & 13.97 & 1.93\\
4 & 40.78\% & 14.15 & 1.94\\
5 & 41.05\% & 14.30 & 1.96\\
6 & 41.07\% & 14.36 & 1.96\\
7 & 41.13\% & 14.36 & 1.96\\
8 & 41.19\% & 14.37 & 1.97\\
9 & 41.26\% & 14.40 & 1.97\\
10 & 41.37\% & 14.37 & 1.98\\
11 & 41.14\% & 14.32 & 1.97\\
12 & 41.19\% & 14.35 & 1.97\\
13 & 40.90\% & 14.31 & 1.96\\
14 & 40.29\% & 14.05 & 1.93\\
15 & 39.81\% & 13.71 & 1.91\\
16 & 38.96\% & 13.35 & 1.88\\
17 & 37.14\% & 12.11 & 1.81\\
18 & 30.96\% & 9.84 & 1.52\\[1ex] 
\hline 
\end{tabular}
\label{table.fixed.pc.bounds} 
\end{table}

\begin{table}[ht]
\caption{Simulation results for the optimized alphas with bounds for $K_1=M$ risk factors ($K_1$ is fixed) but the specific risks ad hoc based on the $K$-factor model ($K$ varies). See Section \ref{sec.conc} for details.} 
\begin{tabular}{l l l l} 
\hline\hline 
$K$ & ROC & SR & CPS\\[0.5ex] 
\hline 
1 & 37.53\% & 11.54 & 1.77\\
2 & 37.93\% & 11.83 & 1.80\\
3 & 38.20\% & 12.03 & 1.82\\
4 & 38.37\% & 12.24 & 1.83\\
5 & 38.70\% & 12.44 & 1.85\\
6 & 38.81\% & 12.56 & 1.86\\
7 & 38.97\% & 12.65 & 1.86\\
8 & 39.14\% & 12.78 & 1.87\\
9 & 39.35\% & 12.91 & 1.88\\
10 & 39.60\% & 13.03 & 1.89\\
11 & 39.58\% & 13.14 & 1.89\\
12 & 39.78\% & 13.27 & 1.91\\
13 & 39.79\% & 13.40 & 1.91\\
14 & 39.39\% & 13.34 & 1.89\\
15 & 39.18\% & 13.20 & 1.88\\
16 & 38.52\% & 13.02 & 1.86\\
17 & 36.99\% & 11.99 & 1.81\\
18 & 30.98\% & 9.85 & 1.52\\[1ex] 
\hline 
\end{tabular}
\label{table.20.pc.xi.bounds} 
\end{table}

\begin{table}[ht]
\caption{Simulation results for the optimized alphas with bounds for varying number $K$ of risk factors and the specific risk ad hoc set to the in-sample risk. See Section \ref{sec.conc} for details. We expect the $K=20$ case to be the same as shrinkage with the shrinkage parameter $q\rightarrow 1$, which is why line 20 in this table is the same (within rounding, which in all tables herein for all non-integer quantities such as ROC, SR, CPS, etc., is to 2 decimal points) as the last line in Table \ref{table.shrinkage}.} 
\begin{tabular}{l l l l} 
\hline\hline 
$K$ & ROC & SR & CPS\\[0.5ex] 
\hline 
1 & 38.80\% & 12.43 & 1.73\\
2 & 39.73\% & 12.96 & 1.77\\
3 & 40.04\% & 13.18 & 1.79\\
4 & 40.25\% & 13.29 & 1.79\\
5 & 40.32\% & 13.38 & 1.80\\
6 & 40.39\% & 13.44 & 1.80\\
7 & 40.43\% & 13.49 & 1.80\\
8 & 40.48\% & 13.53 & 1.80\\
9 & 40.54\% & 13.56 & 1.81\\
10 & 40.58\% & 13.61 & 1.81\\
11 & 40.64\% & 13.63 & 1.81\\
12 & 40.67\% & 13.68 & 1.81\\
13 & 40.63\% & 13.69 & 1.81\\
14 & 40.65\% & 13.72 & 1.81\\
15 & 40.63\% & 13.72 & 1.81\\
16 & 40.65\% & 13.75 & 1.81\\
17 & 40.66\% & 13.77 & 1.81\\
18 & 40.64\% & 13.78 & 1.81\\
19 & 40.65\% & 13.79 & 1.81\\
20 & 40.65\% & 13.81 & 1.81\\[1ex] 
\hline 
\end{tabular}
\label{table.pc.tv.bounds} 
\end{table}

\end{document}